\documentclass[twocolumn,aps,showpacs,floatfix,prc]{revtex4}
\usepackage[dvips]{epsfig}
\begin{document}
\title{ Astrophysical S-factor for the deep sub-barrier fusion reactions of light nuclei }

\author{ Vinay Singh$^{1\S\dagger}$, Debasis Atta$^{2*\dagger}$, Md. A. Khan$^{3**}$ and D. N. Basu$^{4\S\dagger}$}

\affiliation{$^{\S}$Variable Energy Cyclotron Centre, 1/AF Bidhan Nagar, Kolkata 700064, INDIA}
\affiliation{$^{*}$Govt. General Degree College, Kharagpur II, West Bengal 721149, INDIA}
\affiliation{$^{**}$Dept. of Physics, Aliah University, IIA/27, New Town, Kolkata-700156, INDIA}
\affiliation{$^\dagger$Homi Bhabha National Institute, Training School Complex, Anushakti Nagar, Mumbai 400085}

\email[E-mail 1: ]{vsingh@vecc.gov.in}
\email[E-mail 2: ]{debasisa906@gmail.com}
\email[E-mail 3: ]{drakhan.rsm.phys@gmail.com}
\email[E-mail 4: ]{dnb@vecc.gov.in}

\date{\today }

\begin{abstract}

    The cross sections for the deep sub-barrier fusion reaction of light nuclei are calculated within the theoretical framework of the selective resonant tunneling model. In this model, assumption of a complex square-well nuclear potential is invoked to describe the absorption inside the nuclear well. The theoretical estimates for these cross sections agree well with the experimentally measured values. The features of the astrophysical S-factor are derived in terms of this model. Present formalism appears to be particularly useful for the low energy resonant reactions between two charged nuclei.

\vskip 0.2cm


\end{abstract}

\maketitle

\noindent
\section{Introduction}
\label{section1}

    The nuclear reactions play a major role \cite{Bu57,Fo64,Cl83} in determining the structure of main-sequence stars, giant stars, supergiants, pre-supernovae and compact stars like white dwarfs and neutron stars and their evolution and nucleosynthesis they undergo as well as in various observational manifestations. Depending upon the density and temperature along with other parameters, stellar burning is likely to involve many reactions of different kind and involving nuclei from light to heavy and from stable to unstable neutron or proton rich. The rates of these reactions can be calculated from the reaction cross sections $\sigma$ by averaging over a Maxwell-Boltzmann distribution of energies. The Maxwellian-averaged thermonuclear reaction rate $<\sigma v>$ at some temperature $T$, is given by the following integral \cite{Bo08}:

\begin{equation}
 <\sigma v> = \Big[\frac{8}{\pi\mu (k_B T)^3 } \Big]^{1/2} \int \sigma(E) E \exp(-E/k_B T) dE,
\label{seqn1}
\end{equation}
\noindent
where $E$ is the centre-of-mass energy, $v$ is the relative velocity and $\mu$ is the reduced mass of the reactants. At low energies (far below Coulomb barrier) where the classical turning point is much larger than the nuclear radius, barrier penetrability can be approximated by $\exp(-2\pi\zeta)$ so that the charge induced cross section can be decomposed into

\begin{equation}
 \sigma(E) = \frac{S(E)\exp(-2\pi\zeta)}{E}
\label{seqn2}
\end{equation}
\noindent
where $S(E)$ is the astrophysical $S$-factor and $\zeta$ is the Sommerfeld parameter, defined by $\zeta = \frac{Z_1Z_2e^2}{\hbar v}$ where $Z_1$ and $Z_2$ are the charges of the reacting nuclei in units of elementary charge $e$. Except for narrow resonances, the $S$-factor $S(E)$ is a smooth function of energy, which is convenient for extrapolating measured cross sections down to astrophysical energies. In the case of a narrow resonance, the resonant cross section is generally approximated by a Breit-Wigner expression whereas the neutron induced reaction cross sections at low energies can be given by $\sigma(E)=\frac{R(E)}{v}$ \cite{Bl55} facilitating extrapolation of the measured cross sections down to astrophysical energies, where $R(E)$ is a slowly varying function of energy \cite{Mu10} and is similar to $S$-factor.

    The nuclear fusion reactions at very low energies plays important role in nucleosynthesis of light elements in big bang nucleosynthesis as well as inside the stellar core. The fusion cross section is also one of the most important physical quantity required for both design and research in fusion engineering. Nuclear fusion reaction in the energy range of $\sim$1eV to few keV can be explained successfully by the phenomenon of quantum mechanical tunneling through Coulomb barrier of interacting nuclei. In the present work, a simple square-well potential model with an imaginary part has been used to describe the nuclear fusion reaction of light nuclei where the real part of the potential is mainly derived from the resonance energy while the imaginary part is determined by the Gamow factor at resonance energy. This complex square-well nuclear potential describes the absorption inside the nuclear well. The energy dependence of the cross sections and astrophysical S-factors for the deep sub-barrier fusion reactions of light nuclei have been calculated using this model.  

\noindent
\section{Theoretical framework}
\label{section2}

    In case of the light nuclei fusion, treating the resonant tunneling through the Coulomb barrier as a two-step process, that is, first tunneling and then decay, fails to provide an adequate description. Such a oversimplified one-dimensional model \cite{Ga38}, based on the assumption of decay being independent of tunneling, does not depict the true picture of the physical process. In reality, when the wave function of the system of two colliding nuclei  penetrates the barrier, inside the nuclear potential well it reflects back and forth. These reflections inside the nuclear well is completely neglected in the one dimensional model where the wave suffers no reflection while penetrating the barrier. In the case of $\alpha$-decay as well, the outgoing $\alpha$-particle will encounters no reflection after penetrating the barrier in a three dimensional model \cite{Ga28}. While the decay of the penetrating nuclei terminates the bouncing motion inside the nuclear well, if nuclear reaction happens quick enough the wave function will have no time to execute this bouncing motion. In other words, the short lifetime of the penetrating wave may forbid resonant tunneling. This is because of the fact that there will be not enough bouncing motion to built up, inside the nuclear well, the wave function in terms of constructive interference. The tunneling and decay can no longer be independent in light nuclear fusion process and need to be combined as a selective process.

    The lifetime effect on the resonant tunneling can be best achieved by introducing an imaginary part into the nuclear interaction potential. The complex nuclear potential has been shown to describe successfully the resonant tunneling effects in deep sub-barrier fusion using a three dimensional model for wide range of energies \cite{Li99a,Li99b,Li00}. It is precisely, overcoming of the Gamow tunneling insufficiencies by maximizing a damp-matching resonant tunneling. When a light nucleus is injected into another, the reduced radial wave function $\psi(r)$ describing the relative motion can be related to the full wave function $\Phi(r,t)=\frac{1}{\sqrt{4\pi}r}\psi(r)\exp(-i\frac{E} {\hbar}t)$. The full wave function $\Phi(r,t)$ represents the solution of the general Schr\"odinger equation for the system. The reaction cross section in terms of the phase shift $\delta_0$ due to the nuclear potential (in low energy limit only s-wave contributes) is given by $\sigma=\frac{\pi}{k^2}[1- |\eta|^2]$ where $\eta=e^{2i\delta_0}$ and $k=\sqrt{\frac{2\mu E}{\hbar^2}}$ is the wave number for relative motion between the reacting nuclei. In the three dimensional calculation, nuclear potential being complex, the corresponding phase shift $\delta_0$ is complex and is given by \cite{Li00} 
    
\vspace{-0.5cm}  
\begin{eqnarray}
\vspace{-0.25cm}
cot(\delta_0)=W_r+i W_i
\label{seqn3}
\vspace{-0.25cm}
\end{eqnarray}
where instead of conventional phase shift $\delta_0$, a new pair of parameters, $W_r$ and $W_i$, the real and the imaginary parts of the cotangent of phase shift have been introduced. Consequently, the reaction cross section for the s-wave given by $\sigma = \frac{\pi}{k^2}\Big( 1 - |e^{2i\delta_0}|^2\Big)$ can be rewritten as  

\vspace{-0.5cm}
\begin{eqnarray}
\vspace{-0.25cm}
\sigma&=&\frac{\pi}{k^2}\left\{-\frac{4W_i}{(1-W_i)^2+W_r^2}\right\}\\ \nonumber
&=&\left(\frac{\pi}{k^2}\right)\left(\frac{1}{\chi^2}\right)\left\{-\frac{4\omega_i}{\omega_r^2+ (\omega_i-\frac{1}{\chi^2})^2}\right\}
\label{seqn4}
\vspace{-0.25cm}
\end{eqnarray}
\noindent
where $\chi^2$=$\left\{\frac{\exp\left(\frac{2\pi}{ka_c}\right)-1}{2\pi}\right\}$ and $1/\chi^2$ is the Gamow penetration factor, $\omega=\omega_r+i\omega_i=W/\chi^2=(W_r+iW_i)/\chi^2$ and $a_c=\hbar^2/\mu Z_1Z_2e^2$ is the length of Coulomb unit. It is evident that the cross section reaches its maximum when $W_r=0$ and $W_i=-1$ and $W_r=0$ corresponds to the condition for resonance. Thus the condition for resonance is ${\rm Re}(\delta_0) = (2n+1)\pi/2$ where $n$ is an integer. The dimensionless quantity $S_r(E)$ given by

\vspace{-0.5cm}
\begin{eqnarray}
\vspace{-0.25cm}
S_r(E) = \left\{-\frac{4\omega_i}{\omega_r^2+ (\omega_i-\frac{1}{\chi^2})^2}\right\} 
\label{seqn5}
\vspace{-0.25cm}
\end{eqnarray}
\noindent
provides an alternative expression for a dimensionless astrophysical S-function. The wave function inside the nuclear well is determined by two parameters, the real part $V_{r}$ and the imaginary part $V_{i}$ of the nuclear potential. The Coulomb wave function outside the nuclear well is determined by two other parameters: the real and the imaginary part of the complex phase shift $(\delta_0)_r$ and $(\delta_0)_i$. The pair of convenient parameters, $W_r$ and $W_i$, have been introduced to make a linkage between the cross section and the nuclear well so that it is easy to discuss the resonance and the selectivity in damping. The boundary condition for the wave function can be expressed by its logarithmic derivative which for the square well is given by

\vspace{-0.5cm}
\begin{eqnarray}
\vspace{-0.5cm}
R \frac{[\sin(Kr)]^{\prime}}{\sin(Kr)}|_{r=R}=KR\cot(KR)
\label{seqn6}
\vspace{-0.5cm}
\end{eqnarray}
\noindent
and in the Coulomb field, it is given by 

\vspace{-0.5cm}
\begin{eqnarray}
\vspace{-0.25cm}
\frac{R}{a_c}\left\{ \frac{1}{\chi^2} \cot(\delta_0)+ 2\left[ \ln \left(\frac{2R}{a_c}\right)+2A+y(ka_c)\right]\right\}
\label{seqn7}
\vspace{-0.25cm}
\end{eqnarray}
\noindent
so that 

\vspace{-0.5cm}
\begin{eqnarray}
\vspace{-0.25cm}
\omega_i&=&W_i/\chi^2= {\rm Im}\left[ \frac{a_c}{R} (KR)\cot(KR)\right]\\ \nonumber
&=&\frac{a_c}{R} \left\{ \frac{\gamma_i\sin(2\gamma_r)-\gamma_r \sinh  (2\gamma_i)}{2[\sin^2(\gamma_r)+\sinh^2(\gamma_i)]}\right\}\\ \nonumber
\omega_r&=&W_r/\chi^2= \frac{a_c}{R}\left\{  \frac{\gamma_r\sin(2\gamma_r)+\gamma_i \sinh (2\gamma_i)}{2[\sin^2(\gamma_r)+\sinh^2(\gamma_i)]}\right\} \\ \nonumber
&&~~~~~~~~~~~~~~~~-2H \\
\label{seqn8}
\vspace{-0.25cm}
\end{eqnarray}
\noindent
where $K^2=\frac{2\mu}{\hbar^2}[E-(V_r+iV_i)]$, the real part $K_r$ of $K$ and its imaginary part $K_i$ are related by the equation $K_i=\frac{\mu}{K_r\hbar^2}(-V_i)$, $\gamma=(\gamma_r +i \gamma_i)\equiv (K_rR + iK_iR)$,  $H=\left[\ln\left(\frac{2R}{a_c}\right)+2A+y(ka_c)\right]$, radius of the nuclear well $R=r_0(A_1^{1/3}+A_2^{1/3})$, $r_0$ is the radius parameter, $A_1$ and $A_2$ are the mass numbers of the reacting nuclei, Euler's constant $A=0.5772156649$ and and $y(ka_c)$ is related to the logarithmic derivative of $\Gamma$ function given as $y(x)=\frac{1}{x^2}\sum_{j=1}^{\infty} \frac{1}{j(j^2+x^{-2})}-A+\ln(x)$. In the above relations $k=\sqrt{\frac{2\mu E}{\hbar^2}}$ is the wave number outside the nuclear well.

\begin{figure}[t]
\vspace{0.0cm}
\eject\centerline{\epsfig{file=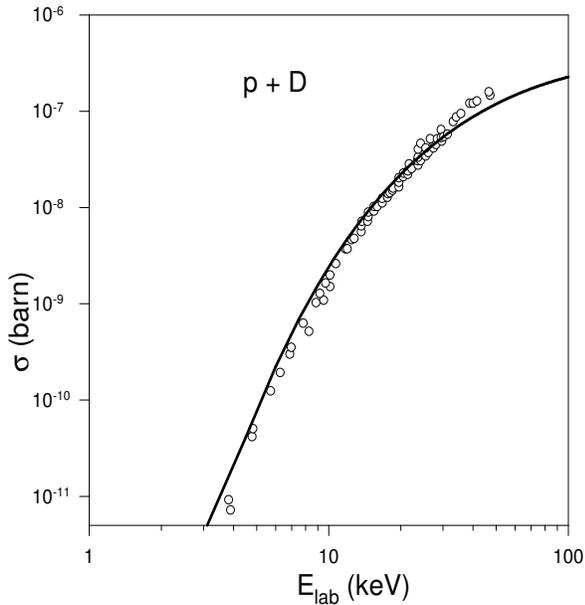,height=8cm,width=7.7cm}}
\caption{Plots of cross-section as a function of lab energy for p+D fusion reaction. The continuous line represents the theoretical calculations while the hollow circles represent the experimental data points.}
\label{fig1}
\vspace{1.0cm}
\end{figure}

\begin{figure}[t]
\vspace{0.0cm}
\eject\centerline{\epsfig{file=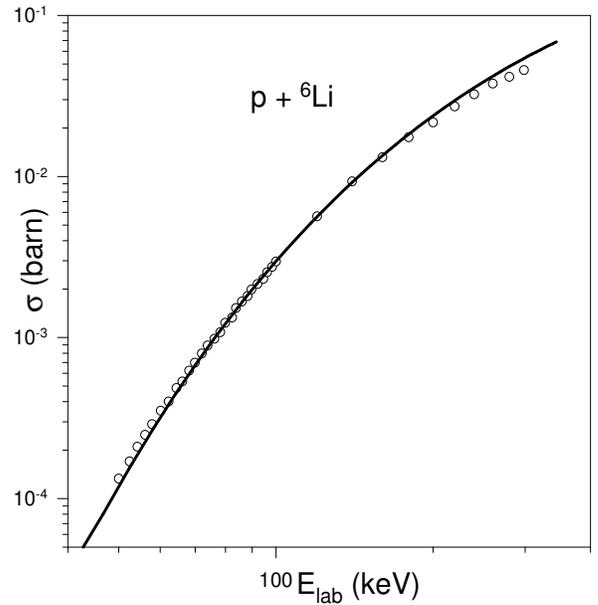,height=8cm,width=7.7cm}}
\caption{Plots of cross-section as a function of lab energy for p+$^6$Li fusion reaction. The continuous line represents the theoretical calculations while the hollow circles represent the experimental data points.}
\label{fig2}
\vspace{1.0cm}
\end{figure}

\noindent
\section{Methodology of theoretical calculations}
\label{section3}

    There are only two adjustable parameters, $V_r$ and $V_i$, in the selective resonant tunneling model. These are adjusted to meet the resonance peak and then it reproduces the data points covering the entire range of energy. The radius parameter $r_0$ may be kept fixed or adjusted to fine tune the calculations. In the present calculations, it is slightly varied from one system to the other in order to obtain better theoretical estimates. The fusion cross sections and the dimensionless astrophysical S-functions are calculated using Eq.(4) and Eq.(5), respectively whereas the astrophysical S-factors (in units of keV-barn) can be calculated using Eq.(4) in Eq.(2). 
   
\noindent
\section{Results and discussion}
\label{section4}

    The present formalism has been used to calculate the fusion cross-sections for D+D, D+T, D+$^3$He, p+D, p+$^6$Li and p+$^7$Li. While the first three \cite{Li99a,Li99b,Li00,Li04,Li08} of these fusion reactions have been done in past with a completely different motive of fusion power production, the rest have been explored in the present work with an intention to use all these six reactions for astrophysical purposes. For the D+D, D+T and D+$^3$He, we use the same $V_r$, $V_i$ and $R$ from Refs. \cite{Li08}, \cite{Li08} and \cite{Li04}, respectively. For the rest of the fusion reactions, $V_r$ and $V_i$ are adjusted to meet the position and magnitude of the resonance peak in the fusion cross-section. The radius parameter $r_0$ (or equivalently the radius of the nuclear well $R$ defined after Eq.(9)) has been further adjusted to fine tune so that the calculations reproduces the experimental data points covering the entire range of energy.

\begin{figure}[t]
\vspace{0.0cm}
\eject\centerline{\epsfig{file=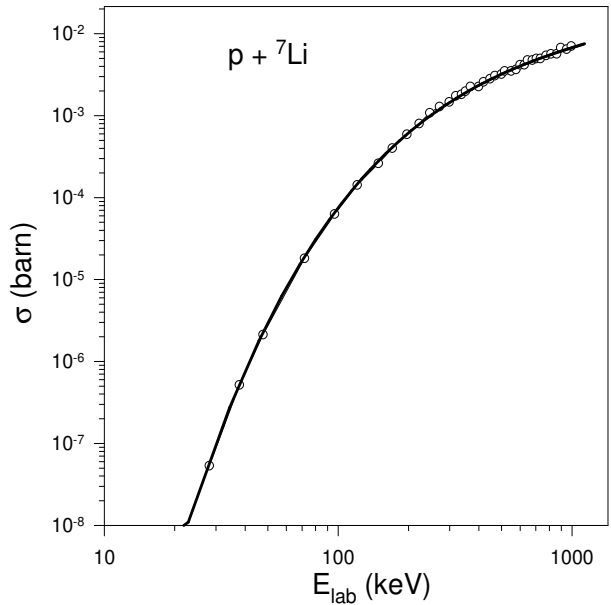,height=8cm,width=8cm}}
\caption{Plots of cross-section as a function of lab energy for p+$^7$Li fusion reaction. The continuous line represents the theoretical calculations while the hollow circles represent the experimental data points.}
\label{fig3}
\vspace{0.0cm}
\end{figure} 
  
\begin{figure}[t]
\vspace{0.0cm}
\eject\centerline{\epsfig{file=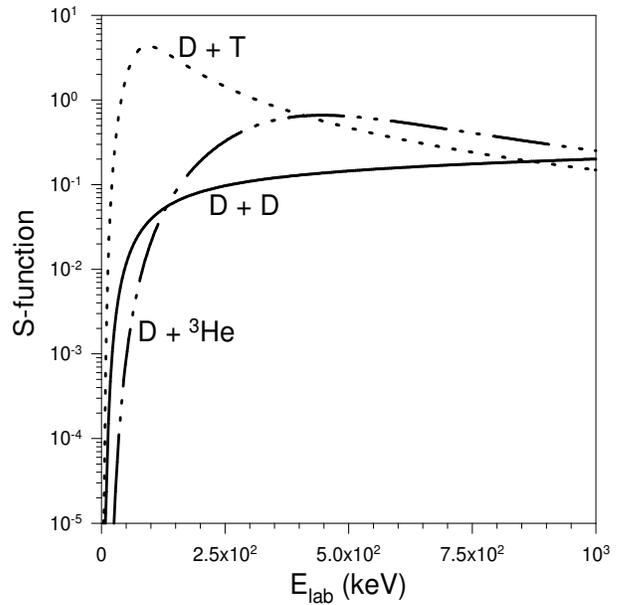,height=8cm,width=8cm}}
\caption{Plots of S-function as a function of lab energy for D+D, D+T, D+$^3$He fusion reactions.}
\label{fig4}
\vspace{1.7cm}
\end{figure} 

\begin{figure}[t]
\vspace{0.0cm}
\eject\centerline{\epsfig{file=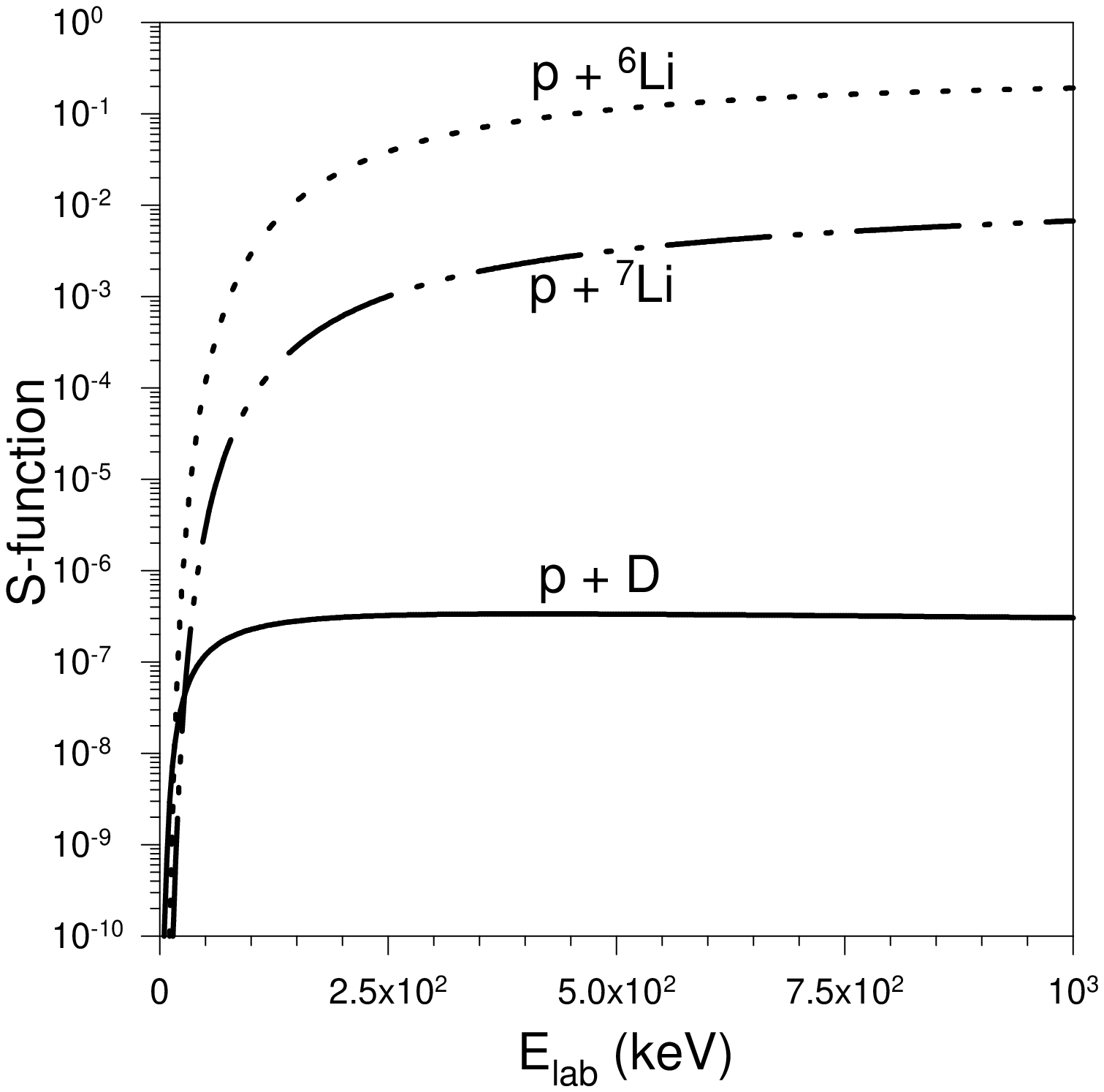,height=8cm,width=8cm}}
\caption{Plots of S-function as a function of lab energy for p+$^{6}$Li, p+$^{7}$Li and p+D fusion reactions.}
\label{fig5}
\vspace{0.0cm}
\end{figure}
     
    The results of the present calculation for cross sections of for D+D, D+T, D+$^3$He fusion reactions have been shown to compare well \cite{Si16} with experimental data as well as those calculated using the three and five parameter fitting formulas of Ref.\cite{Li08}. The results of the cross-section calculations for D+D, D+T, D+$^3$He fusion reactions are available in Ref.\cite{Li08}, Ref.\cite{Li08} and Ref.\cite{Li04}, respectively and the magnitudes of $V_r$, $V_i$ and $r_0$ for these three cases are, respectively, $-48.52$ MeV, $-263.27$ keV, 2.778 fm, $-40.69$ MeV, $-109.18$ keV, 1.887 fm and $-11.859$ MeV, $-259.02$ keV, 3.331 fm. The results of the cross-section calculations for p+D, p+$^6$Li and p+$^7$Li fusion reactions are shown in Figs.1-3, and the magnitudes of  $V_r$, $V_i$ and $r_0$ for these three cases are, respectively, $-55.0$ MeV, $-0.0235$ keV, 1.177 fm, $-44.25$ MeV, $-7.5$ MeV, 1.180 fm and $-85.0$ MeV, $-395.0$ keV, 1.330 fm. It is interesting to note that the magnitude of $V_i$ for fusion of p+$^6$Li is about twenty times larger compared to that of p+$^7$Li. The reason may be attributed to extremely low lifetime of $^8$Be inhibiting its formation. The experimental data \cite{ENDF} and the quantum-mechanical calculations show very good agreement. In similar works, new three parameter formula describes well the low energy behavior of the fusion cross-section for light nuclei \cite{Li12,Liang15}. The results of present calculations for dimensionless S-functions, given by Eq.(5), are shown in Figs.4-5. Somewhat, mismatch with experimental data in case of p+D fusion reaction may be due to lack of experimental data points and any conclusion at this stage regarding drawback of resonance tunneling model in case of heavier nuclei would be improper. However, calculations of fusion cross sections for reactions involving medium and heavy nucleus-nucleus systems do need, altogether, a completely different approach \cite{At14}. 

\noindent
\section{Summary and conclusion}
\label{section5}

    In the deep sub-barrier fusion of light nuclei, the nuclear resonance selects not only the frequency or the energy level but also the damping that causes nuclear reaction. When the resonance occurs, the selectivity becomes very sharp. In such selective resonant tunneling processes the neutron-emission reaction is suppressed. The process of fusion of light nuclei at very low energies can recall the phase factor of the wave function describing the system. The imaginary part in the square well potential describes the formation of compound nucleus \cite{Fe92} formed by the fusion process, but there are not enough collisions to justify the assumptions for compound nucleus model in case of light nuclei. There is no such independent decay process in the light nuclei. In the compound nuclear model, reaction is assumed to proceed in two steps: first fusing to form the compound nucleus followed by its decay. In the present calculations that deal with selective resonant tunneling, the tunneling probability itself depends upon the decay lifetime and is a single step process of fusion of two light nuclei. The agreement with the experimental data for the deep sub-barrier fusion of light nuclei also suggests that the tunneling proceeds in a single step. The recent findings of halo nuclei \cite{Ri94} further strengthens the fact that the nucleons can keep their features without losing memory of the wave function while inside the well of the strongly interacting nuclear region. However, the situation is totally different from the astrophysical reactions, since the weakly bound nature is essential in halo nuclei, whereas a low level density plays an important role in astrophysical reactions implying that the mechanism is different between halo nuclei and astrophysical reactions for a nucleus to retain its identity inside the barrier.

    The complex potential causes absorption of the projectile into the nuclear well. For over last few decades controlled nuclear fusion research has been focused mostly on D-T fusion since it has large fusion cross section compared to that of D-D fusion reaction cross section by a very large factor of the order of several hundred in spite of both having almost the same Coulomb barrier. The resonance of the D+T state near 100 keV is considered as the reason for such a large cross section. A simple square-well potential model with an imaginary part can be used to describe the D+T nuclear fusion as well as other very light nuclei fusion reactions. The D+D, D+T, D+$^3$He, p+$^6$Li, p+$^7$Li and p+D fusion reactions are of astrophysical importance. It is interesting to notice that while the real part of the potential is mainly derived from the resonance energy, the imaginary part of the potential is determined by the Gamow factor at resonance energy. The good agreement between the experimental data and the quantum-mechanical calculation suggests strongly of selective resonant tunneling. The penetrating particle keeps its memory of the phase factor of its wave function. The implication of this selective resonant tunneling model can be further explored for other light nuclei fusion reactions.

\end{document}